\documentstyle[12pt,axodraw]{article}

\advance\textheight by \topskip
\renewcommand{\baselinestretch}{1.2}
\renewcommand{\thefootnote}{\fnsymbol{footnote}}
\newcommand{\beq}{\begin{equation}}
\newcommand{\eeq}{\end{equation}}
\newcommand{\bea}{\begin{eqnarray}}
\newcommand{\eea}{\end{eqnarray}}
\renewcommand{\(}{\left(}
\renewcommand{\)}{\right)}

\renewcommand{\]}{\right]}
\renewcommand{\bar}[1]{\overline{#1}}
\def\slash#1{#1\!\!\!/\!\,\,}
\newcounter{hilf}

\def\SMo{Standard Model}
\def\SM{\SMo\ } \def\SMp{\SMo.\ } 
\def\EW{electro--weak } 
\def\GBo{Goldstone Boson}
\def\GB{\GBo\ }  \def\GBs{\GBo s\ } \def\GBsp{\GBo s.\ }

\begin{document}


\begin{titlepage}
  \renewcommand{\baselinestretch}{1}
  \renewcommand{\thefootnote}{\alph{footnote}}

  \thispagestyle{empty}

   {\bf \hfill                                       MPI--PhT/95--18/rev}

\vspace*{-0.3cm}
   {\bf \hfill                                       TUM--HEP--216/95/rev}

\vspace*{-0.3cm}
   {\bf \hfill                                       LMU--05/95/rev}

\vspace*{-0.3cm}
   {\bf \hfill\hfill                                       July 1995}

\vspace*{0.5cm} {\Large\bf
  \begin{center} Connections between Dynamical and Renormalization Group
                 Techniques in Top Condensation Models
  \end{center}}
\vspace*{1cm}
   {\begin{center} {\large\sc
                Andreas Blumhofer\footnote{\makebox[1.cm]{Email:}
                                  AB@Gluon.HEP.Physik.Uni-Muenchen.DE},
                Richard Dawid\footnote{\makebox[1.cm]{Email:}
                                  Richard.Dawid@Physik.TU-Muenchen.DE}
                and Manfred Lindner\footnote{\makebox[1.cm]{Email:}
                                  Manfred.Lindner@Physik.TU-Muenchen.DE}}
    \end{center} }
\vspace*{0cm} {\it \begin{center}
    \footnotemark[1]Sektion Physik, Ludwig--Maximilians--Universit\"at
    M\"unchen, \\ Theresienstr.37, D--80333 M\"unchen, Germany

                   \

    \footnotemark[2]$\ \!\!^,$\footnotemark[3]Institut f\"ur Theoretische
    Physik, Technische Universit\"at M\"unchen,          \\
    James--Franck--Stra\ss e, D--85747 Garching, Germany
    \end{center} }
\vspace*{2.3cm}

                     {\Large \bf \begin{center} Abstract \end{center} }
                     Predictions for the ratio $M_W/m_t$ arise in top
                     condensation models from different methods. One type of
                     prediction stems from Pagels--Stokar relations based on
                     the use of Ward Identities in the calculation of the \GB
                     decay constants and expresses $M_W$ in terms of integrals
                     containing the dynamically generated mass function
                     $\Sigma_t(p^2)$. Another type of prediction emerges from
                     the renormalization group equations via infrared
                     quasi--fixed--points of the running top quark Yukawa
                     coupling. We demonstrate in this paper that in the limit
                     of a high cutoff these two methods lead to the same
                     predictions for $M_W/m_t$ and $M_W/M_H$ in lowest loop
                     order.

\renewcommand{\baselinestretch}{1.2}
\end{titlepage}

\newpage
\renewcommand{\thefootnote}{\arabic{footnote}}
\setcounter{footnote}{0}


\section{Introduction}

A composite Higgs sector might solve the theoretical problems of scalars
in the \SMp One possibility is that the Higgs particle is essentially a
top--anti-top bound state explaining naturally why the top quark Yukawa
coupling and the Higgs quartic coupling are of order unity (i.e. the largeness
of the top quark and Higgs masses). Instead of fundamental scalars models of
top condensation \cite{BHL,topcon} have therefore some new interaction
capable of forming the required top condensate. The dynamics then generates
an effective scalar sector which describes the symmetry breaking
in analogy to the Ginzburg--Landau
description of superconductivity. The simplest realization of the idea of
top condensation, the so--called BHL model \cite{BHL}, consists just of
the kinetic parts of the ordinary quarks, leptons and
$SU(3)_c\times SU(2)_L\times U(1)_Y$ gauge fields and a new attractive
four--fermion interaction. The BHL Lagrangian is
\beq
{\cal L}_{BHL} = {\cal L}_{kinetic} + G\bar{L}t_R\bar{t}_RL~,
\label{Lbhl}
\eeq
where ${\cal L}_{kinetic}$ contains the kinetic terms for all gauge fields,
quarks and leptons. $L^T=(t_L,b_L)$ is the third generation doublet of
quarks containing the left--handed top and bottom fields and $t_R$ is the
right--handed component of the top quark. The symmetry breaking of the BHL
model by loop effects with a high energy cutoff $\Lambda$ can be studied
in the large $N_c$ limit (where $N_c$ is the number of colors) in analogy
to Nambu--Jona-Lasinio (NJL) models.
For $G> G_{cr} = 8\pi^2/N_c\Lambda^2$ the gap equation is found to be
critical and a top condensate emerges. The breakdown of global symmetries
by this condensate implies the existence of composite \GBsp In the
NJL--treatment of the BHL model these \GBs are found as massless poles in
fermion scattering amplitudes along with a composite Higgs particle of mass
$2m_t$.  It turns out that this set of composite scalar fields is equivalent to
the usual \SM Higgs doublet $\phi$ with a non--vanishing vacuum expectation
value\ (VEV). The symmetry breaking is thus more or less identical to the
\SMp The composite \GBs are therefore eaten exactly as in the \SM and
produce the usual $W$ and $Z$ mass relations.

When the $W$ mass is calculated in large $N_c$ approximation in terms of the
top mass \cite{BHL} then one finds for $N_c=3$
\beq
M_W^2 = m_t^2~\frac{3g_2^2}{32\pi^2}
\ln{\left(\frac{\Lambda^2}{m_t^2}\right)}~.
\label{MWmt}
\eeq
Low values of $\Lambda$ lead to a phenomenologically unacceptable large
top mass and one is thus forced to large $\Lambda$ which requires
however a fine--tuning of $G\rightarrow G_{cr}$. Since the ratio $M_W/m_t$
is so crucial for the phenomenological viability one should study
improvements of the above large $N_c$ relation. Using Ward Identities
one can derive the so--called Pagels--Stokar formulae which can also
be obtained directly by inserting running masses into the derivation
of eq.~(\ref{MWmt}). Another possibility is to use immediately the \SM
as the effective low--energy theory of BHL. This gives constraints on the
running of the \SM top Yukawa coupling $g_t$, which determines the top
mass. In the following we will discuss how these two approaches are
related and we will see that the two methods give the same results in the
high cutoff limit.


\section{Compositeness and the Renormalization Group}

Using the auxiliary field formalism we can rewrite the four--fermion coupling
term of eq.~(\ref{Lbhl}) with the help of a static, non propagating, scalar
doublet $\varphi:=-G\bar{t}_RL$ of mass $G^{-1}$ such that the Lagrangian
eq.~(\ref{Lbhl}) becomes \cite{Eguchi}
\bea
{\cal L}_{aux}= {\cal L}_{kinetic}
               -\bar{L} \varphi t_R-\bar{t}_R \varphi^\dagger L
               -G^{-1} \varphi^\dagger \varphi~.
\label{bhl}
\eea
The dynamics of the original model generates further terms in the Lagrangian
which depend on this scalar field $\varphi$. For large cutoff $\Lambda$ only
renormalizable terms are allowed in the effective Lagrangian such that we
obtain
\beq
{\cal L}_{eff} = {\cal L}_{aux}
       + Z_\varphi \(D_\mu\varphi\)^\dagger
       \(D^\mu\varphi\) + \delta M^2 \varphi^\dagger \varphi -
       \frac{\delta\lambda}{2} \left(\varphi^\dagger \varphi\right)^2 - \delta
       g_t \(\bar{L} \varphi t_R+\bar{t}_R \varphi^\dagger L\)
       +\delta{\cal L}_{kinetic}~.
\label{deltaL}
\eeq
Note that symmetry breaking occurs when $\delta M^2> G^{-1}$ which is achieved
for $G>G_{cr}$. Having the Wilson renormalization group approach in mind we
can immediately read off those conditions which express the composite nature
of the effective scalar Lagrangian
$Z_\varphi \stackrel{p^2\rightarrow\Lambda^2}{\longrightarrow}0$,
$\delta M^2 \stackrel{p^2\rightarrow\Lambda^2}{\longrightarrow}0$, and
$\delta\lambda \stackrel{p^2\rightarrow\Lambda^2}{\longrightarrow}0$.
This expresses simply the fact that all dynamical effects must disappear
as the momentum approaches the scale of new physics.
Additionally we have the normalization conditions
$\delta g_t \stackrel{p^2\rightarrow\Lambda^2}{\longrightarrow}    0$ and
$\delta{\cal L}_{kinetic} \stackrel{p^2\rightarrow\Lambda^2}{\longrightarrow}
0$.
We can now use the freedom to rescale the scalar field  $\varphi$ by defining
$\phi:=\frac{\varphi}{\sqrt{Z_\varphi}}$ such that the Lagrangian
becomes\footnote{We ignore the fermionic wave function contribution as
it does not play any role for the constituent conditions.}
\bea
{\cal L}_{eff} =  {\cal L}_{kinetic}
+ \(D_\mu\phi\)^\dagger \(D^\mu\phi\)
+\frac{\lambda v^2}{2}\phi^\dagger\phi
-\frac{\lambda}{2} \( \phi^\dagger\phi \)^2
-g_t \(\bar{L} \phi t_R+\bar{t}_R \phi^\dagger L\)~.
\label{LSM}
\eea
Here we introduced
$\frac{\lambda v^2}{2} = \frac{\delta M^2 - G^{-1}}{Z_\varphi}$,
$\lambda = \frac{\delta\lambda}{Z_\varphi^2}$ and
$g_t =  \frac{1+\delta g_t}{\sqrt{Z_\varphi}}$ and the effective Lagrangian
has now become the \SMp From the definition of $g_t$, $\lambda$ and $v$ we
see that the conditions of compositeness are
\bea
\lim_{p^2\to\Lambda^2}g_t^{-2}(p^2)= 0~,     \quad
\lim_{p^2\to\Lambda^2}\frac{\lambda(p^2)}{g_t^4(p^2)}= 0~, \quad
\lim_{p^2\to\Lambda^2}\frac{\lambda(p^2) v^2(p^2)}{2g_t^2(p^2)}= -G^{-1}~,
\label{con}
\eea
where $\Lambda$ corresponds to the high energy cutoff of the BHL--model.

These compositeness conditions must obviously be fulfilled in any sensible
non--perturbative treatment of the dynamics like the bubble approximation.
Especially for large $\Lambda$ one may impose the compositeness conditions
directly on the running parameters of the effective Lagrangian (the \SMo).
Conditions~(\ref{con}) lead thus to constraints on the renormalization group
equations of the \SM. Using the full one--loop $\beta$--functions
\cite{betas} these conditions can be studied numerically, but if we restrict
ourself to lowest order $1/N_c$ including the QCD--running we can study the
boundary conditions~(\ref{con}) analytically. The $\beta$--functions
for $N_c =3$ are in this approximation
\bea
\frac{d}{dt}g_t^2 &=& \frac{1}{(4\pi)^2}\(6g_t^2-16g_3^2\)g_t^2
\label{rgt}~, \\
\frac{d}{dt}g_3^2 &=& -\frac{14}{(4\pi)^2}g_3^4~,
\label{rg3}
\eea
and using eq.~(\ref{con}) as boundary condition we find the solutions
\bea
g_t^2(p^2) &=& \frac{g_3^{16/7}(p^2)}
{3\[g_3^{2/7}(p^2)-g_3^{2/7}(\Lambda^2)\]}~,
\label{gt2} \\
g_3^2(p^2) &=&\frac{g_3^2(M_Z^2)}{1+\frac{14g_3^2(M_Z^2)}{(4\pi)^2}t}~,
\label{g32}
\eea
where $t=\frac{1}{2}\ln\frac{p^2}{M_Z^2}$. Thus $m_t=g_t(M_Z^2)v/\sqrt{2}$
with $v=246~GeV$ is predicted by eq.~(\ref{gt2}) in terms of $\Lambda$
and $g_3(\Lambda^2)$. In addition eq.~(\ref{g32}) must be used to
express $g_3^2(\Lambda^2)$ by $\Lambda$ and the known experimental input
$\alpha_3(M_Z^2) = g_3^2(M_Z^2)/4\pi = [0.115-0.125]$.

The predicted top mass is the so called ``infrared quasi--fixed--point''.
This means that the resulting top mass depends for large $\Lambda$ only
extremely mildly on the precise boundary condition at $\Lambda$. This
mild sensitivity can be seen explicitly by demanding
$g_t^{-2}\stackrel{p^2\rightarrow\Lambda^2}{\longrightarrow} \delta$
with e.g. $|\delta|\leq \pi^{-1}$ instead of
$g_t^{-2}\stackrel{p^2\rightarrow\Lambda^2}{\longrightarrow} 0$.
Alternatively one can see that the prediction for the top mass
depends only extremely weakly on $\Lambda$ when $g_3^2(\Lambda^2)$
in eq.~(\ref{gt2}) is expressed by $\alpha_3(M_Z^2)$ via eq.~(\ref{g32}).

The predictions from this renormalization group method should become
more precise as $\Lambda$ becomes larger. The reason is that the
infrared quasi--fixed--point is more attractive for large scales
and that other effects like thresholds etc. should become less
important compared to the renormalization group running. When using this
renormalization group techniques one should however keep in mind that
one assumes quietly that there are no further bound states in the
spectrum\footnote{Or equivalently that all other states have mass
$\Lambda$ such that they do not contribute to the running.}.


\section{Predictions from Pagels--Stokar Formulae}

The dynamically generated $W$--mass is determined by the eaten composite
Goldstone modes. The \GB decay constants are essentially given by the
fermion--loop self energies of the condensing fermion while the other
contributions are strongly suppressed. Using Ward Identities the
$W$--mass can be calculated in terms of the momentum dependent mass function
of the top quark $\Sigma_t(p^2)$ with Euclidean momentum $p$. This leads
to the well known Pagels--Stokar formula \cite{PAGELS,BLULI}
\bea
    M_W^2 = \frac{3 g_2^2}{2(4\pi)^2} \int\limits_0^{\Lambda^2} dp^2
            \frac{\Sigma_t^2(p^2)}{p^2+\Sigma_t^2(p^2)}  ~,
\label{Pagels}
\eea
where $\Sigma_t(p^2)$ is the solution of the full gap (Schwinger--Dyson)
equation.The top mass is defined at an Euclidean scale $M_Z$ as
\bea
      m_t= \Sigma_t(M_Z^2)~.
\label{onshellt}
\eea
It is possible to understand the origin of the
dynamically generated mass function $\Sigma_t(p^2)$
in different pictures. First $\Sigma_t(p^2)$ can be
viewed as the solution of a Schwinger--Dyson equation
of the underlying theory (e.~g. Topcolor or another renormalizable model
\cite{Hill}).
Alternatively $\Sigma_t(p^2)$ can be described in terms
of the effective Lagrangian plus additional ``residual''
effects of the underlying theory. Such residual effects
should however show up only at $p^2 = {\cal O}(\Lambda^2)$.
This is a consequence of the decoupling theorems in renormalizable
theories and a necessary condition for interpreting the \SM as an
effective theory of top condensation.
For $p^2 \ll \Lambda^2$ the mass function $\Sigma_t(p^2)$
should therefore be described very well by the effective
Lagrangian.

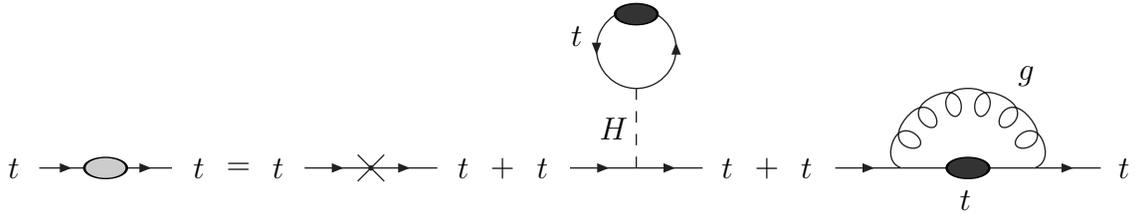
\begin{figure}[htb]
\begin{center}
\begin{picture}(320,80)(-50,10)
\ArrowLine(-80,10)(-60,10)        \ArrowLine(-50,10)(-30,10)
\GOval(-55,10)(4,8)(0){0.8}
\Text(-90,10)[c]{$t$}            \Text(-20,10)[c]{$t$}
\Text(-5,10)[c]{=}
\ArrowLine(20,10)(45,10)        \ArrowLine(45,10)(70,10)
\Line(40,5)(50,15)              \Line(50,5)(40,15)
\Vertex(45,10) 1
\Text(10,10)[c]{$t$}            \Text(80,10)[c]{$t$}
\Text(95,10)[c]{+}
\ArrowLine(120,10)(145,10)      \ArrowLine(145,10)(170,10)
\DashLine(145,10)(145,40)4
\ArrowArc(145,55)(15,270,90)    \ArrowArc(145,55)(15,90,270)
\GOval(145,68)(4,8)(0){0.2}
\Text(110,10)[c]{$t$}             \Text(180,10)[c]{$t$}
\Text(137,25)[c]{$H$}             \Text(123,60)[c]{$t$}
\Text(195,10)[c]{+}
\ArrowLine(220,10)(245,10)      \ArrowLine(245,10)(295,10)
\ArrowLine(295,10)(320,10)
\GlueArc(270,10)(25,0,180)5 6
\GOval(270,10)(4,8)(0){0.2}
\Text(210,10)[c]{$t$}           \Text(330,10)[c]{$t$}
\Text(290,45)[l]{$g$} \Text(270,-2)[c]{$t$}
\end{picture}
\end{center}
\caption{\it \SM Schwinger--Dyson equation for the top mass including
QCD corrections. The grey dot denotes the top self energy $\Sigma_t(p^2)$
and full dots are top quark propagators $(\slash{p}-\Sigma_t(p^2))^{-1}$.
The bare mass is symbolized by a cross.}
\label{SDSM}
\end{figure}

\begin{figure}[htb]
\begin{center}
\begin{picture}(310,40)(0,10)
\ArrowLine(20,10)(40,10)        \ArrowLine(50,10)(70,10)
\GOval(45,10)(4,8)(0){0.8}
\Text(10,10)[c]{$t$}            \Text(80,10)[c]{$t$}
\Text(95,10)[c]{=}
\ArrowLine(120,10)(145,10)      \ArrowLine(145,10)(170,10)
\Vertex(145,10) 2
\ArrowArc(145,25)(15,270,90)    \ArrowArc(145,25)(15,90,270)
\GOval(145,38)(4,8)(0){0.2}
\Text(110,10)[c]{$t$}             \Text(180,10)[c]{$t$}
\Text(123,30)[c]{$t$}
\Text(195,10)[c]{+}
\ArrowLine(220,10)(245,10)      \ArrowLine(245,10)(295,10)
\ArrowLine(295,10)(320,10)
\GlueArc(270,10)(25,0,180)5 6
\GOval(270,10)(4,8)(0){0.2}
\Text(210,10)[c]{$t$}           \Text(330,10)[c]{$t$}
\Text(290,45)[l]{$g$} \Text(270,-2)[c]{$t$}
\end{picture}
\end{center}
\caption{\it Top condensation gap equation in the same approximation as
the case of the \SMp}
\label{SDTC}
\end{figure}
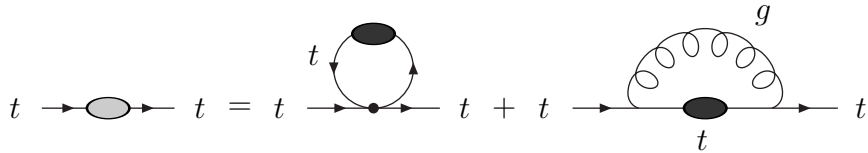

In the framework of gap equations
one can see this by understanding the full \SM top mass function as the
solution of a Schwinger--Dyson equation (see fig.~\ref{SDSM}).
For simplicity we include only one loop QCD--corrections.
In solving this system one regains the renormalization group running
of $m_t$. Now we compare this equation with the corresponding top condensation
Schwinger--Dyson equation (see fig.~\ref{SDTC}).
The residual effects of the former discussion
correspond to the difference between the effective four--fermion coupling
in fig.~\ref{SDTC} and a more complicated interaction structure. These heavy
degrees of freedom can be integrated out below the heavy boson mass scale.
The equations in fig.~\ref{SDSM} and fig.~\ref{SDTC} can be identified
in one regularization scheme by fitting the bare mass in fig.~\ref{SDSM}
appropriately.
The same kind of identification can also be done including scalars after
a reshuffling of the perturbation series in the top condensate
model (see exact renormalization group method e.g. \cite{Polch} ).
Thus in the high cutoff limit the solution of the full gap equation
$\Sigma_t(p^2)$ fulfills the standard model renormalization group equation.

Now, interpreting the \SM as an effective description of a top condensation
model, we identify the masses with their \SM definitions. For the $W$--mass we
have $M_W={g_2v}/{2}$, the running of $\Sigma_t$ should be connected to the
corresponding renormalization group of the effective top--Yukawa coupling.
A naive approach for the top mass function therefore leads to
\bea
       \Sigma_t(p^2)=\frac{g_t(p^2)v}{\sqrt{2}}~,
\eea
and we get from the following constituent condition from eq.~(\ref{Pagels}):
\bea
       1  = \frac{3}{(4\pi)^2} \int\limits_0^{\Lambda^2} dp^2
            \frac{g_t^2(p^2)}{p^2+g_t^2(p^2)\frac{v^2}{2}}~.
\label{gap1}
\eea

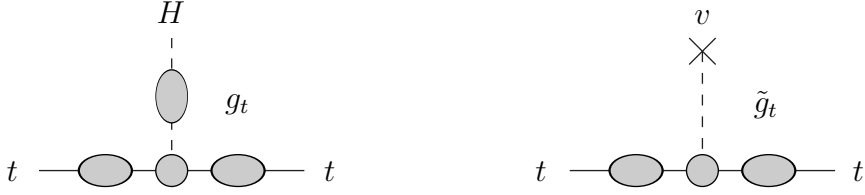
\begin{figure}[htb]
\begin{center}
\begin{picture}(340,80)(0,0)
\Line(20,10)(120,10)
\DashLine(70,10)(70,60)4
\GCirc(70,10)6 {0.8}
\GOval(45,10)(6,10)(0){0.8}
\GOval(95,10)(6,10)(0){0.8}
\GOval(70,38)(10,6)(0){0.8}
\Text(10,10)[c]{$t$}            \Text(130,10)[c]{$t$}
\Text(95,35)[c]{$g_t$}          \Text(70,70)[c]{$H$}
\Line(220,10)(320,10)
\DashLine(270,10)(270,55)4
\GCirc(270,10)6 {0.8}
\GOval(245,10)(6,10)(0){0.8}
\GOval(295,10)(6,10)(0){0.8}
\Text(210,10)[c]{$t$}           \Text(330,10)[c]{$t$}
\Text(295,35)[c]{$\tilde{g}_t$}         \Text(271,68)[c]{$v$}
\Line(265,50)(275,60)           \Line(265,60)(275,50)
\end{picture}
\end{center}
\caption{\it Quantum corrections to $g_t$ and $\tilde{g}_t$.}
\label{gtgt}
\end{figure}

For large $\Lambda$ we can neglect the low energy behaviour of $g_t(p^2)$
and replace $\Sigma_t$ in the denominator by an infrared cutoff of the \EW
scale:
\bea
    \frac{3}{(4\pi)^2} \int\limits_{M_Z^2}^{\Lambda^2}
    \frac{g_t^2(p^2)}{p^2} dp^2 = 1~.
    \label{ig1}
\eea
If the running of $g_t$ is fixed by the renormalization group then
eq.~(\ref{ig1}) constitutes a top mass prediction since it can be
used to determine the single free boundary condition. As both the
infrared quasi--fixed--point prediction eq.~(\ref{gt2}) and our new
prediction eq.~(\ref{ig1}) should become more precise for large $\Lambda$
one should expect that both results are consistent or at least
approximately consistent. But inserting eqs.~(\ref{gt2}) and (\ref{g32}) into
eq.~(\ref{ig1}) leads to
\bea
    \frac{3}{(4\pi)^2} \int\limits_{M_Z^2}^{\Lambda^2}
    \frac{g_t^2(p^2)}{p^2} dp^2 =
    \frac{3}{7} \int\limits^{g_3^2(M_Z^2)}_{g_3^2(\Lambda^2)}
    g_t^2 \frac{dg_3^2}{g_3^4}  \to  \infty~,
    \label{inf}
\eea
i.e. this consistency of the two predictions appears badly violated.
The solution to this problem can be found in the scale dependence of
$\Sigma_t$. The renormalization group equation describes the change of
the couplings corresponding to a simultaneous rescaling  of all the outer
momenta. But the contribution to $g_t$ which stems from the Higgs line
cannot contribute to $\Sigma_t$ because there is no momentum flow into
the Higgs leg which ends in the vacuum. To describe this situation correctly
we introduce a new running coupling constant $\tilde{g}_t$ which is defined
by
\bea
       \Sigma_t = \tilde{g}_t \frac{v}{\sqrt{2}}
       \label{gttilde}~,
\eea
and does not contain wave function corrections of the Higgs (see
fig.~\ref{gtgt}).
\begin{figure}[htb]
\begin{center}
\begin{picture}(250,200)
\epsfysize=8.0cm
\epsffile[600 100 1100 600]{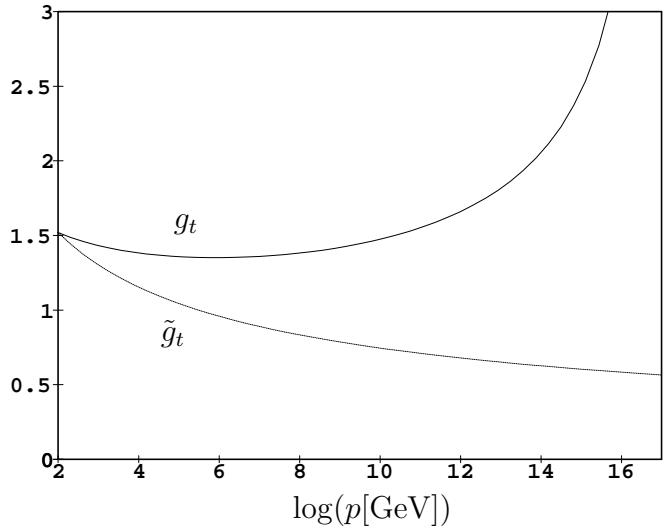}
\end{picture}
\end{center}
\vspace*{-8cm}
\begin{center}
\begin{picture}(250,200)
\put(45,100){\mbox{$g_t$}}
\put(40,57){\mbox{$\tilde{g}_t$}}
\put(90,-10){\mbox{$\log(p[\rm{GeV}])$}}
\end{picture}
\end{center}
\caption{\it $g_t$ and $\tilde{g}_t$ for $g_3^2(M_Z^2)=1.52$
             and $\Lambda=10^{17}{\rm GeV}$.}
\label{tm}
\end{figure}
The connection between  $g_t$ and $\tilde{g_t}$ can be seen more clearly
if we keep in mind that $g_t$ is a three point function with two independent
momenta. Let ${\rm g}_t(p^2,q^2)$ be the Yukawa coupling for
parallel\footnote{Angles between $p$ and $q$ will not be important for our
purposes.} incoming top momentum $p$ and Higgs momentum $q$.Then $g_t$ and
$\tilde{g_t}$ describe the renormalization group running with respect to
different momentum arguments:
\bea
       g_t(p^2) &=& {\rm g}_t(p^2,p^2)          ~, \label{gtpp}\\
       \tilde{g}_t(p^2) &=& {\rm g}_t(p^2,M_Z^2)~. \label{gtpMZ}
\label{pq}
\eea
This naturally implies $\tilde{g}_t(M_Z^2)=g_t(M_Z^2)$ such that the wave
function of the Higgs field is normalized to one at the scale of the Higgs
mass.

The renormalization group equation describes the running of $g_t$.
Eq.~(\ref{rgt}) may in fact be derived by calculating
\beq
\frac{d}{dt} g_t^2 =
\sigma \frac{d}{d\sigma}~ {\rm g}_t^2((\sigma p)^2, (\sigma p)^2)
\label{betasigma}
\eeq
from the relevant diagrams. Similarly we may formally define a corresponding
renormalization group equation for the newly defined quantity $\tilde{g}_t$
and calculate the corresponding $\beta$--function in large $N_c$--limit
\beq
        \frac{d}{dt} \tilde{g}_t^2 =
        \sigma \frac{d}{d\sigma}~ \tilde{\rm g}_t^2((\sigma p)^2, M_Z^2) =
        \frac{1}{(4\pi)^2}\(-16g_3^2\)\tilde{g}_t^2~.
        \label{tgt}
\eeq
We solve eq.~(\ref{tgt}) using the boundary condition
$\tilde{g}_t(M_Z^2)=g_t(M_Z^2)$ with $g_t(M_Z^2)$ defined by eq.~(\ref{gt2})
and obtain
\bea
    \tilde{g}_t^2(p^2) &=& \frac{g_3^{16/7}(p^2)}
    {3\[g_3^{2/7}(M_Z^2)-g_3^{2/7}(\Lambda^2)\]}~.
    \label{tgt2}
\eea
If we insert this result instead of $g_t$ into the left hand side of
eq.~(\ref{ig1}) we find
\bea
    \frac{3}{(4\pi)^2} \int\limits_{M_Z^2}^{\Lambda^2}
    \frac{\tilde{g}_t^2(p^2)}{p^2} dp^2  =
    \frac{3}{7} \int\limits^{g_3^2(M_Z^2)}_{g_3^2(\Lambda^2)}
    \tilde{g}_t^2 \frac{dg_3^2}{g_3^4}  = 1~.
    \label{rel1}
\eea
Hence the Pagels--Stokar condition is fulfilled if we use correctly
$\tilde{g}_t$ instead of $g_t$. Eq.~(\ref{ig1}) must be written therefore
correctly as
\bea
    \frac{3}{(4\pi)^2} \int\limits_{M_Z^2}^{\Lambda^2}
    \frac{\tilde{g}_t^2(p^2)}{p^2} dp^2 = 1~,
    \label{ig1t}
\eea
and the Pagels--Stokar formula is then in perfect agreement with the
corresponding renormalization group running.

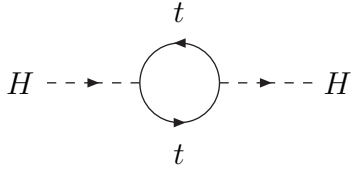
\begin{figure}[htb]
\begin{center}
\begin{picture}(150,60)(0,-10)
\DashArrowLine(15,20)(50,20)4   \DashArrowLine(80,20)(115,20)4
\ArrowArc(65,20)(15,0,180)      \ArrowArc(65,20)(15,180,360)
\Text(5,20)[c]{$H$}            \Text(125,20)[c]{$H$}
\Text(65,47)[c]{$t$}            \Text(65,-7)[c]{$t$}
\end{picture}
\end{center}
\caption{\it Higgs self energy diagram.}
\label{hsed}
\end{figure}

There is a nice way to see this consistency of eq.~(\ref{ig1t}) without
explicitly inserting eq.~(\ref{tgt2}). Consider the evolution of
$\tilde{g}_t^2 / g_t^2$:
\bea
      \frac{d}{dt}\frac{\tilde{g}_t^2}{g_t^2} =
      \(\frac{\frac{d}{dt}\tilde{g}_t^2}{\tilde{g}_t^2}-
      \frac{\frac{d}{dt}g_t^2}{g_t^2}\)\frac{\tilde{g}_t^2}{g_t^2}~.
      \label{ba}
\eea
If we insert then eq.~(\ref{rgt}) and eq.~(\ref{tgt}) into eq.~(\ref{ba}),
we obtain immediately
\bea
      \frac{d}{dt}\frac{\tilde{g}_t^2}{g_t^2} =
      -\frac{6}{(4\pi)^2} \tilde{g}_t^2~.
      \label{dg}
\eea
According to fig.~(\ref{gtgt}) this equation involves only the wave function
correction of the Higgs sector and at one loop level this is just the single
diagram in fig.~(\ref{hsed}).

Integrating eq.~(\ref{dg}) leads to:
\bea
      \left.\frac{\tilde{g}_t^2}{g_t^2}\right|^{\Lambda^2}_{M_Z^2} =
      -\frac{3}{(4\pi)^2}
      \int\limits^{\Lambda^2}_{M_Z^2} \tilde{g}_t^2 \frac{dp^2}{p^2}~.
\eea
Since the compositeness condition requires a pole for $g_t$ while
eq.~(\ref{tgt}) implies that $\tilde{g}_t$ tends to zero we find
that $\tilde{g}_t^2/g_t^2$ goes to zero at $\Lambda$. Together with
the normalization condition $\tilde{g}_t(M_Z)=g_t(M_Z)$ one obtains
therefore exactly the result eq.~(\ref{rel1}) without making use of
explicit solutions of the renormalization group equations. In this way
we can even add other terms or new interactions which contribute to
$g_t$ and $\tilde{g}_t$ but do not change the Higgs wave function.

It is interesting that the $W$--mass generated by a single top--loop
is consistent with the renormalization group method using only
the graph in fig. (\ref{hsed}). This can be understood by considering
that the $W$--mass is connected to the \GB wave function by
gauge symmetry. Since custodial $SU(2)$ symmetry is just
broken by finite terms, the logarithmically divergent pieces of the
\GB wave function and the Higgs wave function must be identical.
Only these logarithmically divergent terms contribute to the renormalization
group equation.


\section{The Quartic Higgs Coupling}

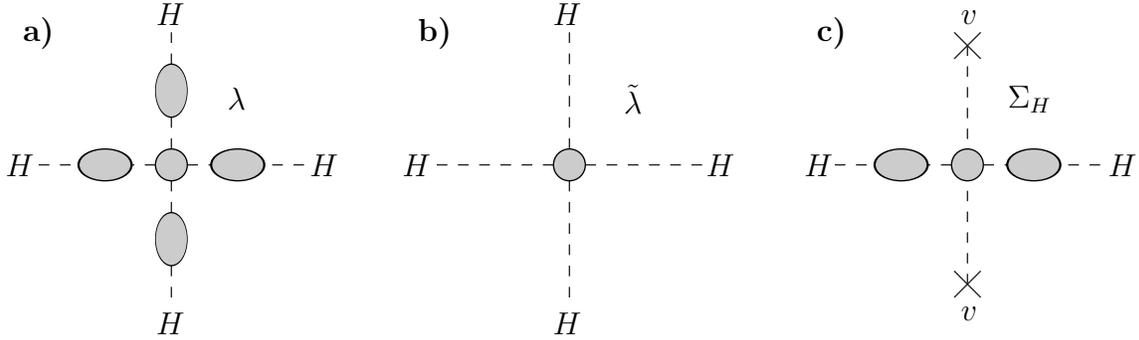
\begin{figure}[htb]
\begin{center}
\begin{picture}(430,120)(0,-60)
\DashLine(10,10)(110,10)4
\DashLine(60,-40)(60,60)4
\GCirc(60,10)6 {0.8}
\GOval(35,10)(6,10)(0){0.8}
\GOval(85,10)(6,10)(0){0.8}
\GOval(60,38)(10,6)(0){0.8}
\GOval(60,-18)(10,6)(0){0.8}
\Text(3,10)[c]{$H$}              \Text(118,10)[c]{$H$}
\Text(60,-50)[c]{$H$}            \Text(60,67)[c]{$H$}
\Text(10,60)[c]{\bf a)}          \Text(85,35)[c]{$\lambda$}
\DashLine(160,10)(260,10)4
\DashLine(210,-40)(210,60)4
\GCirc(210,10)6 {0.8}
\Text(153,10)[c]{$H$}            \Text(268,10)[c]{$H$}
\Text(210,-50)[c]{$H$}           \Text(210,67)[c]{$H$}
\Text(160,60)[c]{\bf b)}         \Text(235,35)[c]{$\tilde{\lambda}$}
\DashLine(310,10)(410,10)4
\DashLine(360,-35)(360,55)4
\GCirc(360,10)6 {0.8}
\GOval(335,10)(6,10)(0){0.8}
\GOval(385,10)(6,10)(0){0.8}
\Text(305,10)[c]{$H$}            \Text(420,10)[c]{$H$}
\Text(362,-46)[c]{$v$}           \Text(362,66)[c]{$v$}
\Line(355,50)(365,60)            \Line(355,60)(365,50)
\Line(355,-40)(365,-30)          \Line(355,-30)(365,-40)
\Text(310,60)[c]{\bf c)}         \Text(385,35)[c]{$\Sigma_H$}
\end{picture}
\end{center}
\caption{\it Quantum corrections to $\lambda$, $\tilde{\lambda}$ and
             $\Sigma_H$.}
\label{ldad}
\end{figure}

The techniques of the last sections can also be applied to the Higgs mass
and the quartic Higgs coupling $\lambda$. The renormalization group
equation for $\lambda$ is in large $N_c$ approximation
\bea
    \frac{d}{dt}\lambda = \frac{12}{(4\pi)^2}g_t^2(\lambda-g_t^2)~.
\eea
Note that the term proportional to $\lambda^2$ is missing because it
is suppressed by $1/N_c$. Using the compositeness conditions eq.~(\ref{con})
and the above result for $g_t$ the running $\lambda$ is
\bea
    \lambda(p^2) = \frac{2\[g_3^{18/7}(p^2)-g_3^{18/7}(\Lambda^2)\]}
                        {27\[g_3^{2/7}(p^2)-g_3^{2/7}(\Lambda^2)\]^2}~.
\eea
As in the case of the top Yukawa coupling we define a coupling
$\tilde{\lambda}$ which does not contain any Higgs wave function
corrections (see fig.~(\ref{ldad})):
\bea
    \tilde{\lambda} = \lambda \frac{\tilde{g}_t^4}{g_t^4}~,
\eea
so that the renormalization group equation for $\tilde{\lambda}$ reads
\bea
    \frac{d}{dt} \tilde{\lambda} = -\frac{12}{(4\pi)^2} \tilde{g}_t^4~.
\eea
\vspace{.5cm}

We find therefore
\bea
    \tilde{\lambda}(M_Z^2) = \left.-\lambda \frac{\tilde{g}_t^4}{g_t^4}
                   \right|^{\Lambda^2}_{M_Z^2}
    = \int\limits^{\Lambda^2}_{M_Z^2}
      \frac{d}{dt}(-\tilde{\lambda})\frac{dp^2}{2p^2}
    = \frac{6}{(4\pi)^2}\int\limits^{\Lambda^2}_{M_Z^2}\tilde{g}_t^4
      \frac{dp^2}{p^2}~.
    \label{lda}
\eea
Now we have to introduce the \SM Higgs mass definition. The Higgs mass
function $\Sigma_H(p^2)$ which corresponds to fig.~(\ref{ldad}c) involves
twice the coupling $g_t$ and twice $\tilde{g}_t$. Expressed in $\lambda$
this means:
\bea
    \Sigma_H^2(p^2)= \lambda \frac{\tilde{g}_t^2(p^2)}{g_t^2(p^2)}v^2~.
\eea
With our normalization condition $g_t(M_Z^2)=\tilde{g}_t(M_Z^2)$
insertion of this mass function into eq.~(\ref{lda}) leads to:
\bea
    M_H^2 = \Sigma_H^2(M_Z^2) =
    \frac{12}{(4\pi)^2}\int\limits^{\Lambda^2}_{M_Z^2}
    \tilde{g}_t^2 \Sigma_t^2(p^2) \frac{dp^2}{p^2}~.
    \label{mh}
\eea
This is a corresponding Pagels--Stokar formula for $M_H$ which was also
found by V.~N.~Gribov \cite{GRIBOV} for a specific case.
One can directly evaluate this formula in analogy to the Pagels--Stokar
calculations of the $W$--mass using the top--loop diagram with four outer
Higgs lines. The outer momenta are kept at the infrared cutoff so that the
Yukawa couplings can be replaced by the running $\tilde{g}_t$ to improve the
diagram. In this way one arrives at eq.~(\ref{mh}).

We have demonstrated that the renormalization group equations lead to
the same result for $M_W$ and $M_H$ as the direct calculation via
Pagels--Stokar formulae. Moreover the behavior of $\lambda$ and $g_t$
at the scale $\Lambda$ shows the real power of the renormalization
group analysis for top condensation models in the following calculation.
The ratio of the running Higgs and top mass squared is defined by
\bea
   \frac{\Sigma_H^2(p^2)}{\Sigma_t^2(p^2)}=
   \frac{\lambda(p^2) \frac{\tilde{g}_t^2(p^2)}{g_t^2(p^2)}v^2}
   {\tilde{g}_t^2(p^2) \frac{v^2}{2}}
   = \frac{2\lambda(p^2)}{g_t^2(p^2)}
   = \frac{4\[g_3^{18/7}(p^2)-g_3^{18/7}(\Lambda^2)\]}
          {9g_3^{16/7}(p^2)\[g_3^{2/7}(p^2)-g_3^{2/7}(\Lambda^2)\]}~.
\eea
One can easily find the square root of this ratio in the limit
$p^2\to\Lambda^2$:
\bea
   \lim_{p^2\to\Lambda^2} \frac{\Sigma_H(p^2)}{\Sigma_t(p^2)} = 2~,
\eea
which is precisely what we expect: The binding energy goes to zero at
the condensation scale so that the bound states do not exist above
$\Lambda$. This result requires the $1/N_c$--expansion and does not
hold in the one--loop \SMo, which does not respect that expansion.


\section{Conclusions}

We studied in this paper connections between two techniques which
predict mass ratios $m_t/M_W$ and $M_H/M_W$ in top condensation models
with an effective scalar sector. One method arises from infrared
quasi--fixed--points in the renormalization group running of the
effective Yukawa coupling, the other from using Pagels--Stokar relations
which are based on Ward Identities. We argue that for a
high cutoff it should be possible to identify the top mass function in
Pagels--Stokar just with the running mass function of the
standard model. Nevertheless the insertion of the
running \SM top Yukawa coupling $g_t$ in the Pagels--Stokar formula
leads to an infinite $W$--mass. This apparent
contradiction between our two approaches comes from the fact that
the top mass function $\Sigma_t$ of the \SM is not controlled by the
running coupling $g_t$, but by a modified running coupling $\tilde{g}_t$.
Taking this into account we show at one loop that in the high cutoff limit
the Pagels--Stokar relation and the renormalization group method are
equivalent. This equivalence can be used to construct an analogous
Pagels--Stokar formula for the Higgs mass by imposing corresponding boundary
conditions on the renormalization group flow of the quartic Higgs
coupling $\lambda$. Recently this relation was obtained by Gribov
\cite{GRIBOV} in a specific model.
Additionally we find the expected mass relation $\Sigma_H = 2 \Sigma_t$ at the
compositeness scale $\Lambda$.

Since the techniques which are applied in this paper are rather general
we believe that our results are much more generally valid in
composite Higgs models with effective Yukawa and Higgs couplings along the
lines of reasoning after eq.~12.

\vspace{.5cm}
{\bf Acknowledgments:} We would like to thank E. Schnapka for useful comments
on the draft version of this paper.
This work is in part supported by DFG and DAAD.



\vspace{1.cm}

\begin{thebibliography}{50}
\bibitem{BHL}    W.A.~Bardeen, C.T.~Hill and M.~Lindner, Phys.~Rev.~D41
                 (1990) 1647.
\bibitem{topcon} Y. Nambu, in {\it New Theories in Physics},
                 proceedings of the XI International Symposium on Elementary
                 Particle Physics, edited by Z.~Ajduk, S.~Pokorski and
                 A.~Trautman (World Scientific, Singapore, 1989);\\
                 A. Miransky, M. Tanabashi and K. Yamawaki,
                 Mod. Phys. Lett. A4 (1989) 1043, Phys. Lett. B221 (1989) 177.
\bibitem{Eguchi} T.~Eguchi, Phys.~Rev.~D14 (1976) 2755;\\
                 F.~Cooper, G.~Guralnik and N.~Snyderman, Phys.~Rev.~Lett.~40
                 (1978) 1620.
\bibitem{betas}  M.E. Machacek and M.T. Vaughn, Nucl.~Phys.~B222 (1983) 83;
                 Nucl.~Phys.~B236 (1984) 221; Nucl.~Phys. B249 (1985) 70.
\bibitem{PAGELS} H.~Pagels and S.~Stokar, Phys.~Rev.~D20 (1979) 2947; \\
                 A.~Carter and H.~Pagels, Phys.~Rev.~Lett.~43 (1979) 1845;\\
                 R.~Jackiw and K.~Johnson, Phys.~Rev.~D8 (1973) 2386.
\bibitem{BLULI}  Similar relations can also be obtained by different methods:\\
                 A.~Blumhofer and M.~Lindner, Nucl.~Phys.~B407 (1993) 173.
\bibitem{Hill}   S.~King and S.~Mannan, Phys.~Lett.~B241 (1990) 249; \\
                 C.~T.~Hill, Phys.~Lett.~B266 (1991) 419; \\
                 R.~B\"onisch, Phys.~Lett.~B268 (1991) 394; \\
                 M.~Lindner and D.~Ross, Nucl.~Phys.~B370 (1992) 30.
\bibitem{Polch}  J.~Polchinski, Nucl.~Phys.~B231 (1984) 269.
\bibitem{GRIBOV} V.~N.~Gribov, Phys.~Lett.~B336 (1994) 243.
\end{thebibliography}
\end{document}